# White noise jammer mathematical modelling and simulation

**Theodoros G. Kostis**

**Dept of Electronics, Electric Power & Telecommunications, Hellenic Air Force Academy, Dekeleia Air Force Base, Aharnes 13671, TGA1010, Greece**
tkostis@aegean.gr

**Abstract:** The radar equation is the fundamental mathematical modelling of the basic function of a radar system. Moreover there are many versions of the radar equation that correspond to particular radar operations, like a low PRF, a high PRF or a surveillance mode. In many cases all these expressions of the radar equation exist in their combined forms giving little information to the actual physics and signal geometry between the radar and target involved in the process. In this case study we divide the radar equation into its major steps and present a descriptive mathematical modelling of the radar and other related equations utilizing the free space loss and the target gain concepts for simulating the effect of a white noise jammer to an adversary radar. We believe that this work will be particularly beneficial to instructors of radar courses and to radar simulation engineers because of its analytical block approach of the main equations related to radar and electronic warfare fields. Finally this work falls under the field of predictive dynamics for radar systems using mathematical modelling techniques.

## 1. Introduction

Usually the radar equation is presented in the literature in its combined form. The disadvantage of this depiction is that there are a lot of physics in the process that remain hidden. Furthermore in the afore-mentioned combined form the signal geometry of the scene between the radar and the target is not shown in all its important stages. Of course the analytical form of the radar equation, with all its various steps, exists in almost all books on radar engineering [1], [2], [3], [4],[5], [6], [7], [8].

Our contribution in this case study is that we manipulate the mathematical formula of the radar equation to come up with a term called gain of the target, which of course includes the radar cross section σ. The advantage of using this target gain term is that it tries to explain the target's reflectivity by treating the target as an antenna gain. Of course this value can be negative or positive in decibel notation according to the reflectivity type of target involved.

In chapter 2 we employ the telecommunications equation in order to set up the foundations of the numerical analysis for the radar and electronic equations in the rest of chapters.

Exactly with this equation we can proceed to derive the radar equation in chapter 3. Also here we coin the aforementioned target gain parameter.

In chapter 4 we use the noise of the system and communications channel in order to bring out the important concept of Signal-to-Noise Ratio (SNR).

In chapter 5 we make the parameter values of the generic telecommunications equation more specific in order to explain the Radar Warning Receiver (RWR) equation and in chapter 6 we use the same methodology in order to set up the jammer equation and the important electronic warfare term of Jammer-to-Signal Ratio (JSR).

Finally the decibel notation is mentioned throughout this work because it makes the task of radar calculations and simulations extremely easy in comparison with real numbers.

## 2. Employing the Telecommunications Equation

The Friis Transmission Equation describes the budget link between a transmitter and a receiver. We will employ this concept in order to derive the mathematical models for telecommunications, radar and electronic warfare fields.

First we begin with the power that is created by the emitter of a telecommunications subsystem.

$$P_{TX}\ [W] \xrightarrow{dB} [dBW]$$
$$P_{TX}\ [mW] \xrightarrow{dB} [dBm]$$
$$dBm = dBW + 30 \quad (1)$$

where:

$P_{TX}$ is the transmitted power

Now this power is translated to electromagnetic waves by the antenna subsystem.

When the radiation is perfectly uniform in a three-dimensional space, it is called isotropic.

$$G = 1 \xrightarrow{dB} G_{dB} = 0\,dB \quad (2)$$

In the case of more directive antennas, the gain can be expressed by considering the solid angle of their radiation.

$$G = \eta_{ant} \frac{4\pi}{\theta_{az}\theta_{el}} [\text{no units}] \Leftrightarrow$$
$$G = \eta_{ant} \frac{4\pi}{\theta_{az}\theta_{el}} \Leftrightarrow \eta_{ant} \frac{4\pi}{\Omega_{beam}} \left[\frac{rad^2}{rad^2}\right] \quad (3)$$

where:

$\eta_{ant}$ is the antenna efficiency

$\theta_{az}$ is the azimuth angle of the radar beam

$\theta_{el}$ is the elevation angle of the radar beam

$\Omega_{beam}$ is the beam solid angle of the radar

Now the Effective Isotropic Radiated Power, which is the combination of the emitter transmission power and the gain of the antenna, is the final power emitted by the communications system.

$$EIRP = P_{TX} G_{TX} \qquad (4)$$

where:

$P_{TX}$ is the transmitted power

$G_{TX}$ is the radar's antenna gain towards the other station's receiver, or to a radar target or generally towards a surveillance sector.

This power is divided by the surface of a sphere, because of the spherical nature of electromagnetic waves.

$$S_{TX} = P_{TX} G_{TX} \frac{1}{4\pi R^2} \qquad (5)$$

where:

$S_{TX}$ is the transmitted power density that meets the receiver's antenna

$R$ is the distance between the transmitter and the receiver

Now the amount of energy that the receiver's antenna can process is dependent upon the antenna effective area $A_{eff|RX}$.

$$P_{RX} = P_{TX} G_{TX} \frac{1}{4\pi R^2} A_{eff|RX} \qquad (6)$$

where:

$A_{eff|RX}$ is the receiver's effective antenna area

The antenna's effective area is dependent upon the gain of the antenna multiplied by a frequency and geometrical factor. Moreover this formula is derived by equating the blackbody spectral radiance of the radio waves to the Johnson-Nyquist noise power of the antenna resistance, which leads to equation (7)

$$A_{eff} = \frac{\lambda^2}{4\pi} \xrightarrow{G_{RX} \neq 1} A_{eff} = \frac{\lambda^2 G_{RX}}{4\pi} \qquad (7)$$

Informationally we can also express the reception of an antenna in terms of a solid angle $\Omega$.

$$\Omega = \frac{A_{eff}}{R^2} = \frac{\frac{G_{RX} \lambda^2}{4\pi}}{\frac{R^2}{1}} = \frac{\lambda^2}{4\pi R^2} G_{RX} \qquad (8)$$

The above information helps to bring the telecommunications equation to a more geometrically oriented form

$$P_{RX} = P_{TX} G_{TX} \frac{1}{4\pi R^2} \frac{\lambda^2 G_{RX}}{4\pi}$$
$$P_{RX} = P_{TX} G_{TX} \frac{\lambda^2}{(4\pi R)^2} G_{RX} \qquad (9)$$

Using the factor L which is the free space loss from the Friis equation several telecommunications links can be modelled in an effective manner using this convenient loss parameter, as shown in Figure 1.

$$L = \frac{(4\pi R)^2}{\lambda^2} \Leftrightarrow \frac{1}{L} = \frac{\lambda^2}{(4\pi R)^2} \qquad (10)$$

where :

$L$ is the free space loss from the Friis equation

Moreover the free space loss is further calculated, by using the dB notation, with the following forms which either include the wavelength or the frequency of the telecommunications system.

$$L = 20\log(4\pi) + 20\log R_{meters} - 20\log \lambda_{meters}$$
$$L = +21.98 dB + 20\log R_{meters} - 20\log \lambda_{meters}$$
$$L = +92.45 dB + 20\log R_{Km} + 20\log f_{GHz} \qquad (11)$$
$$L = +32.45 dB + 20\log R_{Km} + 20\log f_{MHz}$$

Further on we arrive at the final form of the telecommunications equation, in real numbers, which shows that the received power is the product of the transmitted power, the transmitter antenna's gain, the free space loss and the receiver antenna's gain.

$$P_{RX} = P_{TX} G_{TX} \frac{1}{L} G_{RX} \qquad (12)$$

Now expressing (12) in decibel units, multiplication becomes addition and division becomes subtraction producing the final form of the telecommunications equation in dB units.

$$P_{RX} = P_{TX} + G_{TX} + (-L) + G_{RX} \qquad (13)$$

Of course all the above math agree with the Friis equation, which is as follows:

$$\frac{P_{RX}}{P_{TX}} = \frac{A_{effTX} A_{effRX}}{\lambda^2 R^2} = \frac{\frac{\lambda^2 G_{TX}}{4\pi} \frac{\lambda^2 G_{RX}}{4\pi}}{\lambda^2 R^2} \qquad (14)$$

$$P_{RX} = P_{TX} G_{TX} \frac{\lambda^2}{(4\pi R)^2} G_{RX} \qquad (15)$$

### 3. Switching to the Radar Equation

The telecommunications equation for one way path can be altered to represent the radar equation which includes two way paths.

Using (5) we add the rest of the radar path and come up with a power term instead of a power density term. Analytically the next step is the illumination of the target by the radar transmitter antenna, which is backscattered by the target's radar cross section $\sigma$.

$$P_{RX|FROM\_TARGET} = P_{TX} G_{TX} \frac{1}{4\pi R^2} \sigma \quad (16)$$

The energy that is returned by the target to the radar is again distributed in the surface of a sphere.

$$P_{RX|BEFORE\_ANTENNA} = P_{TX} G_{TX} \frac{1}{4\pi R^2} \sigma \frac{1}{4\pi R^2} \quad (17)$$

The final step is the target's returned energy that meets the radar's reception antenna effective area.

$$P_{RX} = P_{TX} G_{TX} \frac{1}{4\pi R^2} \sigma \frac{1}{4\pi R^2} A_{eff}$$
$$P_{RX} = P_{TX} G_{TX} \frac{1}{4\pi R^2} \sigma \frac{1}{4\pi R^2} \frac{\lambda^2 G_{RX}}{4\pi} \quad (18)$$

Now by employing some mathematics we will express the radar equation:

$$P_{RX} = P_{TX} G_{TX} \frac{\lambda^2}{(4\pi R)^2} \sigma \frac{1}{4\pi R^2} G_{RX} \quad (19)$$

By inserting the necessary factors that will create the return free space loss, we have

$$P_{RX} = P_{TX} G_{TX} \frac{\lambda^2}{(4\pi R)^2} \sigma \left[\frac{4\pi}{\lambda^2}\right] \frac{\lambda^2}{(4\pi R)^2} G_{RX}$$
$$P_{RX} = P_{TX} G_{TX} \frac{\lambda^2}{(4\pi R)^2} \left[\sigma \frac{4\pi}{\lambda^2}\right] \frac{\lambda^2}{(4\pi R)^2} G_{RX} \quad (20)$$

At this point we have coined the parameter that represents the gain of the target $G_{TRG}$ as a radiation parameter.

$$G_{TRG} = \frac{4\pi\sigma}{\lambda^2}$$
$$G_{TRG|dB} = 10\log(4\pi) + 10\log\sigma - 20\log\lambda \quad (21)$$

Further on we treat the target as an antenna gain, which is dimensionless, that radiates backscattered energy to the radar, like an antenna in a steradian.

An explanation comes handy from [7] which states that the radar cross section can be derived to be dimensionless and [8] which explains the radar cross section theory, as in (22).

$$\sigma = 4\pi \frac{\Phi_{scattered}}{S_{incident}} = \left[\frac{\frac{W}{srad}}{\frac{W}{m^2}}\right] = [m^2] \quad (22)$$

Therefore,

$$G_{TRG} = \frac{4\pi}{\lambda^2}\sigma = \left[\frac{4\pi}{m^2} m^2\right] = [\text{no units}] \quad (23)$$

This notation may be proven useful in modelling longer wavelength radar links, which do not travel in a straight line but use multiple hops, as a single dimensionless parameter.

Now the radar equation becomes:

$$P_{RX} = P_{TX} G_{TX} \frac{\lambda^2}{(4\pi R)^2} G_{TRG} \frac{\lambda^2}{(4\pi R)^2} G_{RX} \quad (24)$$

Of course when the above equation is collapsed it is the same as the usual radar equation that is found in all literature

$$P_{RX} = \frac{P_{TX} G_{TX} G_{RX} \sigma \lambda^2}{(4\pi)^3 R^4} \quad (25)$$

The next step is to express equation (24) from real numbers notation to decibel values.

$$P_{RX} = P_{TX} + G_{TX} + (-L) + G_{TRG} + (-L) + G_{RX} \quad (26)$$

### 4. Thermal Noise, Channel Noise and Signal-to-Noise Ratio (SNR)

There are three major factors that degrade the received signal: thermal noise, receiver noise factor and the channel's additive white noise.

#### 4.1. Thermal Noise

In any temperature above absolute zero (-273.15C or 0 Kelvin) the friction of the atoms in the electronics of the receiver will produce white noise or Johnson Noise, in dB.

$$N_{Johnson|dB} = N_{thermal|dB} = k_{B|dB} + T_{0|dB} + B_{dB} \quad (27)$$

When we add the losses from the receiver's amplifiers or circuits, which are represented by the receiver's Noise Factor $NF$ we come up with:

$$N_{RX|dB} = k_{B|dB} + T_{0|dB} + B_{dB} + NF_{dB} \quad (28)$$

### 4.2. Channel Noise

We simulate the channel's additive white noise by introducing the MOD factor. An explanation is that the sensitivity of the receiver is worsened by channel's noise. Moreover the channels noise is considered to be white noise, therefore very similar to the thermal noise in the receiver.

In other words the modulation used needs to achieve a higher SNR for successful demodulation. This approach agrees with the fact that channel's noise can be seen as degrading the sensitivity of the receiver.

Therefore the total noise that sets the minimum discernible signal of the receiver is as follows:

$$N_{TOTAL|dB} = k_{B|dB} + T_{0|dB} + B_{dB} + NF_{dB} + MOD_{dB} \quad (29)$$

### 4.3. Signal-to-Noise Ratio (SNR)

An important metric that exhibits the quality of signal reception is the Signal-to-Noise Ratio, which is as follows:

$$SNR = \frac{P_{RX}}{N_{TOTAL}} \text{ or}$$
$$SNR = P_{RX} - N_{TOTAL} \text{ in dB} \quad (30)$$

The SNR metric is particularly important in radar since it is directly connected with the probability of detection and probability of false alarm concepts.

### 5. Making variations: Radar Warning Receiver Equation

A radar warning receiver is a system that detects the radio emissions of radar systems and issues a warning when such signals might be a threat to the carrying platform.

The mathematical modelling of an RWR can be now made very simple by using the afore-mentioned methodology using the decibel notation.

$$P_{RWR|RX} = P_{RDR|TX} + G_{RDR|TX} + (-L) + G_{RWR|RX} \quad (31)$$

where:

$P_{RWR|RX}$ is the power of the radar locking onto the platform and which the RWR can see

We could attempt the same SNR analysis for the RWR, where the noise of radar receiver is given by the usual noise mathematical formula.

$$N_{RWRTOTAL} = k_B + T_0 + B + NF + MOD \quad (32)$$

Further on we could define the SNR of the radar warning receiver as:

$$SNR_{RWR} = P_{RWRRX} - N_{RWRTOTAL} \quad (33)$$

But usually we measure the performance of the RWR by considering only its sensitivity $P_{RWR|RX}$ and not taking into account the respective SNR. For example for the detection of pulsed radars the sensitivity of the RWR should be at least -40dBm and for CW signals the corresponding sensitivity should be -50dBm [1].

Of course more complex simulation of advanced radar warning receivers that try to detect low probability waveform is outside the scope of the current case study.

### 6. More variations: Jammer Equation and Jammer-to-Signal Ratio (JSR)

Another transformation of the telecommunications equation allows for the mathematical modelling of a jammer to a radar system.

#### 6.1. Jamming-to-Signal Ratio (JSR)

The effect that a conventional jammer has on a radar is the increase of the white noise in the receiver, thus degrading the SNR. In this manner the detection probability of a target is also degraded and the jammer has achieved its purpose. The jammer equation, with all units in dB is shown below:

$$J = P_{JAM|TX} + G_{JAM|TX} + (-L) + \frac{B_{RDR}}{B_{JAM}} + G_{RDR|RX} \quad (34)$$

where:

$J$ is the jammer power that meets the radar receiver's antenna

$P_{JAM|TX}$ is the transmitted power from the jammer

$G_{JAM|TX}$ is the jammer's antenna gain

$B_{RDR}$ is the bandwidth of the radar's receiver

$B_{JAM}$ is the bandwidth of the jammer

$G_{RDR|RX}$ is the radar's antenna gain in reception towards the jammer

The term $\frac{B_{RDR}}{B_{JAM}}$ means that if the jammer's bandwidth is bigger that the radar's bandwidth, like as in barrage jamming, then the advantage goes to the radar. On the other hand, if the jammer can pinpoint the bandwidth of the radar, then the advantage goes to the jammer.

On another note because the overall white noise of the radar receiver is quite small compared to the white noise created by the jammer, it is usually omitted. Here the term jammer-to-Signal Ratio (JSR) is coined by the electronic warfare community to demonstrate the effectiveness of the jammer's effect, which is (in dB)

$$JSR = (N_{TOTAL} + J) - P_{RX} = J - P_{RX} \quad (35)$$

Of course in order to calculate the new SNR at the radar after the jamming effect, the conjugate concept of the JSR, which is the Signal-to-Jammer Ratio (SJR) is utilized.

Therefore the new SNR of the radar, after the jammer's effect, is (in dB)

$$SJR = P_{RX} - (N_{TOTAL} + J)$$
$$N_{TOTAL} << J \quad (36)$$
$$SJR_{N_{TOTAL}<<J} = P_{RX} - J$$

## 7. SJR & JSR Simulation

In decoy mode an electronic attack jammer must acquire the operational characteristics of the adversary radar so it can inject false targets. On the other hand there is no need to acquire the adversary radar's operating details when the jamming function is using white noise. In other words the major effort of a white noise jammer is the degradation of the adversary radar's SNR.

For example let's simulate, using Mathcad™, the invasion of a typical fighter or radar cross section of $3m^2$, a stealthier target like a ballistic missile or a B-2 with $\sigma=1m^2$ and a much stealthier target like an F-35 which is claimed to be at $\sigma=0.005\ m^2$, starting at 250Km to selected ranges up to 10Km away from an early warning radar, like the AESA [9] AN/FPS-117. This type of early earning radar operates at 24.6kW peak power and 1.3GHz operating frequency. It's radar's antenna is set to provide +40.59dBi at azimuth of 0.18 degrees and full elevation of 20 degrees (fan shaped) and the bandwidth is set to 100MHz to provide for pulse compression issues. Moreover an additional noise from the radar's amplifiers (noise factor equals 5dB was added and a typical further atmospheric noise (MOD=10dB) was used.

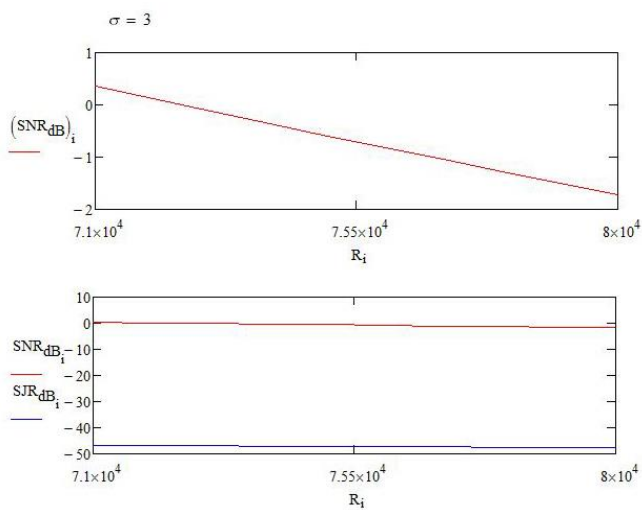

*Fig.1 Simulator graphical results*

As far as the jammer is concerned, we could set the jammer emitter power to 6800W, like the AN/ALQ-99E, its antenna gain to +22.63dBi and its jamming bandwidth to 1GHz.

Then, as shown in Table 1, we took a sample of results from the simulator which corresponds to ranges of interest for air-defence reasons.

Additionally the simulator provides a graphical representation of the results, as can be seen in Figure 1 which represents such a sample.

From the results of Figure 2 can see that without the jammer a target will eventually be seen by the radar, but with the jammer the target will be seen much later by the radar after it passes the burnthrough range. Moreover the modular layout of the simulator is shown in Figure 2.

## 8. Discussion

This paper is an exercise in electronic warfare, it tries to predict the effect of a white noise jammer to a radar system and therefore in the radar science falls under the category of predictive dynamics.

As W. F. Bahret stated in an interview for the "Cold War Technology History Project", (2006) [10], there are three ways to calculate a radar discipline, in the interview's case the radar echo, in our paper's case the performance of a white noise jammer. One is to measure the actual machine. And the third is to use scale models or of course actual real setups. But these methods require that somebody took the risk of creating something first which cost time and funding that bears a probability that may not work well at the end.

Bahret's way which this paper adheres to is the second option, mathematical techniques, that is "techniques that permit you to design something and predict the outcome" [10, page 6]. Moreover mathematical techniques which take the form of predictive dynamics, is a very heavily practice in the radar discipline because of all the probabilities of intercept, detection and false alarm that are involved together with a considerable amount of partial information that exists between a radar and a target.

Therefore we back up our claims using the predictive dynamics of mathematical modelling & techniques which is appropriate for the radar and electronic warfare fields.

Analytically we started with the telecommunications equation as a foundation stone in order to derive the radar equation. In this process the useful concept of $G_{TRG}$ was shown which models the backscattering of a target as a dimensionless antenna gain. Furthermore this modelling angle may be proven useful for the simulation of all propagation situations, including multi-hop HF radar.

Furthermore the well-established concept of SNR for radar detection was shown particularly in its decibel form, which provides for easier computations.

Then we conveniently altered the value of the parameters in the telecommunications equation in order to come up with the simulation of a Radar Warning Receiver and a Jammer. Moreover the electronic warfare concepts of JSR and SJR were depicted.

Further additions to the above equations that describe any kind of losses, like antenna polarization mismatches and radar antenna impedance mismatches, were intentionally omitted in this case study for clarity of results.

## 9. Conclusion

In this paper we arrived at the radar equation by using the telecommunications equation. Moreover we discussed the target gain $G_{TRG}$ which gives an intuitive variation of the backscattering of a target as an antenna gain in a solid angle. This information was used to create a prediction dynamics tool for assessing the performance of a white noise jammer using mathematical modelling techniques. Finally in this case study several well-established concepts in radar engineering and electronic warfare fields were depicted in their convenient for calculations decibel form.

### Telecommunications Link

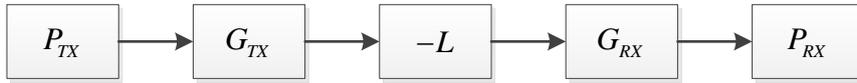

### Radar

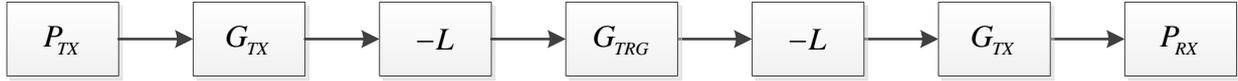

### Radar Warning Receiver

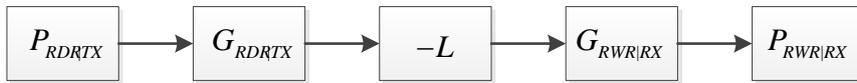

### Jammer

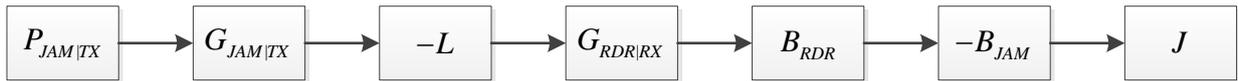

*Fig. 2.* Modular mathematical modelling of several telecommunications links using the free space loss block

|  | Stealthier Target | | Stealth Target | | Fighter type | |
|---|---|---|---|---|---|---|
|  | F-35 | | Ballistic Missile, B-2 | | Mig-21 | |
|  | SNR | SJR | SNR | SJR | SNR | SJR |
|  | $\sigma=0.005[m^2]$ | | $\sigma=0.1[m^2]$ | | $\sigma=3[m^2]$ | |
| 10.00 | (+4.97dB) | (-58.27dB) | (+17.98dB) | (-45.26dB) | (+32.75dB) | (+30.49dB) |
| 12.00 | (+2.07dB) | (-59.73dB) | (+15.08dB) | (-46.71dB) | (+29.85dB) | (+31.94dB) |
| **13.60** | **(+0.05dB)** | **(-60.73dB)** | (+13.06dB) | (-47.72dB) | (+27.83dB) | (-32.95dB) |
| 15.00 | (-2.50dB) | (-61.53dB) | (+11.47dB) | (-48.52dB) | (+26.24dB) | (-33.75dB) |
| 20.00 | (-6.00dB) | (-63.89dB) | (+6.75dB) | (-50.88dB) | (+21.52dB) | (-36.11dB) |
| **29.00** | (-12.45dB) | (-66.99dB) | **(+0.50dB)** | **(-53.98dB)** | (+15.32dB) | (-39.21dB) |
| 30.00 | (-13.00dB) | (-67.27dB) | (-0.01dB) | (-54.26dB) | (+14.75dB) | (-39.49dB) |
| **70.00** | (-27.42dB) | (-74.47dB) | (-14.41dB) | (-61.46dB) | **(+0.30dB)** | **(-46.69dB)** |
| 100.00 | (-33.50dB) | (-77.53dB) | (-20.53dB) | (-64.52dB) | (-5.76dB) | (-49.75dB) |
| 250.00 | (-49.40dB) | (-85.44dB) | (-36.34dB) | (-72.43dB) | (-21.57dB) | (-57.66dB) |

*Table. 1.* Comparison of SNR and respective SJR for three types of targets